\documentclass[twocolumn]{aastex63}

\usepackage{mathrsfs}
\usepackage[hyphens]{url}

\received{2020 June 16}
\revised{2020 September 14}
\accepted{2020 September 19}
\published{2020 October 27}

\shorttitle{LEO Satellite Trail Mitigation}
\shortauthors{Tyson et al.}

\begin{document}

\title{Mitigation of LEO Satellite Brightness and Trail Effects on the Rubin Observatory LSST}


\author[0000-0002-9242-8797]{J. Anthony Tyson}
\affiliation{Department of Physics, University of California, One Shields Ave., Davis, CA  95616, USA; \textup{\url{tyson@physics.ucdavis.edu}}}
\affiliation{Rubin Observatory, Tuscon, AZ 85719, USA}

\author[0000-0001-5250-2633]{\v{Z}eljko Ivezi\'{c}}
\affiliation{Department of Astronomy, University of Washington, Seattle, WA  98195, USA}
\affiliation{DIRAC Institute, University of Washington, Seattle, WA  98195, USA}
\affiliation{Rubin Observatory, Tuscon, AZ 85719, USA}

\author[0000-0002-8998-6739]{Andrew Bradshaw}
\affiliation{SLAC, 2575 Sand Hill Rd, Menlo Park, CA 94025, USA}

\author[0000-0003-1305-7308]{Meredith L. Rawls}
\affiliation{Department of Astronomy, University of Washington, Seattle, WA  98195, USA}
\affiliation{DIRAC Institute, University of Washington, Seattle, WA  98195, USA}

\author[0000-0002-9171-4046]{Bo Xin}
\affiliation{Rubin Observatory, Tuscon, AZ 85719, USA}

\author[0000-0003-2874-6464]{Peter Yoachim}
\affiliation{Department of Astronomy, University of Washington, Seattle, WA  98195, USA}
\affiliation{DIRAC Institute, University of Washington, Seattle, WA  98195, USA}

\author{John Parejko}
\affiliation{Department of Astronomy, University of Washington, Seattle, WA  98195, USA}
\affiliation{DIRAC Institute, University of Washington, Seattle, WA  98195, USA}

\author{Jared Greene}
\affiliation{SpaceX, 22908 NE Alder Crest Dr., Redmond, WA, USA}

\author[0000-0002-3583-4692]{Michael Sholl}
\affiliation{SpaceX, One Rocket Rd., Hawthorn, CA 90250, USA}

\author[0000-0003-1587-3931]{Timothy M. C. Abbott}
\affiliation{NOIRLab, MSO/CTIO, La Serena, Chile}

\author[0000-0001-7445-4724]{Daniel Polin}
\affiliation{Department of Physics, University of California, One Shields Ave., Davis, CA  95616, USA; \textup{\url{tyson@physics.ucdavis.edu}}}

\begin{abstract}
We report studies on the mitigation of optical effects of bright low-Earth-orbit (LEO) satellites on Vera C. Rubin Observatory and its Legacy Survey of Space and Time (LSST). These include options for pointing the telescope to avoid satellites, laboratory investigations of bright trails on the Rubin Observatory LSST camera sensors, algorithms for correcting image artifacts caused by bright trails, experiments on darkening SpaceX Starlink satellites, and ground-based follow-up observations.
The original Starlink v0.9 satellites are $g \sim$ 4.5 mag, and the initial experiment ``DarkSat'' is $g\sim6.1$ mag. Future Starlink darkening plans may reach $g\sim7$ mag, a brightness level that enables nonlinear image artifact correction to well below background noise.
However, the satellite trails will still exist at a signal-to-noise ratio $\sim100$, generating systematic errors that may impact data analysis and limit some science. For the Rubin Observatory 8.4-m mirror and a satellite at 550 km, the full width at half maximum of the trail is about $3^{\prime\prime}$ as the result of an out-of-focus effect, which helps avoid saturation by decreasing the peak surface brightness of the trail.
For 48,000 LEOsats of apparent magnitude 4.5, about 1\% of pixels in LSST nautical twilight images would need to be masked.

\vspace{0.5em}
\noindent\textit{Unified Astronomy Thesaurus concepts:} Artificial satellites (68); CCD observation (207); CCD photometry (208); Observational astronomy (1145); Astronomical techniques (1684); Photometry (1234); Astronomy data analysis (1858); Astronomy data acquisition (1860); Astronomy data reduction (1861); Field of view (534); Sky
surveys (1464) \\
\end{abstract}


\section{Introduction} \label{sec:intro}

\setcounter{footnote}{8}

Innovation in spacecraft manufacturing and launch technology has resulted in a profusion of proposals to build, launch, and operate constellations of many low-Earth-orbit\footnote{For the purposes of this paper, we apply the Low Earth Orbit definition of satellites in a “spherical region that extends from the Earth’s surface up to an altitude (Z) of 2,000 km,” as identified in the Space Debris Mitigation Guidelines of the Inter-Agency Debris Coordination Committee (IADC) and the United Nation’s Office Of Outer Space Affairs (UNOOSA).} (LEO) commercial satellites.  Currently, about one thousand operational LEO satellites (LEOsats) provide communications and earth-imagery services, but regulatory applications filed with international agencies project an increase by {over 100 fold} in the next 5--10 yr.  Many such constellations are either U.S. licensed or have sought permission to operate in the U.S. There are also several other LEOsat operators in other countries with plans to launch their own constellations\footnote{\url{https://en.wikipedia.org/wiki/Satellite\_constellation}}. Several LEOsat projects plan to offer global broadband services. In order to offer low-latency internet access to less-populated areas of the world, companies are proposing constellations of unprecedented size. While it is unclear how many of the proposed LEOsat projects will receive funding to build and deploy, the prospect of $>48,000$ LEOsats in aggregate would represent a potentially significant impact for optical astronomy.

LEOsats scatter sunlight for several hours after sunset or before sunrise, are relatively close to Earth and bright, and can affect ground-based optical observations
\citep{ 2020A&A...636A.121H, 2020ApJ...892L..36M, 2020AAS...23541003S}. The impact of individual LEOsats on astronomy depends on the rate of interfering luminous trails and their brightness, which are in turn affected by spacecraft design and their operational attitude.
Both of these factors are exacerbated for large wide-field ground-based facilities such as Vera C. Rubin Observatory and its planned 10-year Legacy Survey of Space and Time (LSST).

Ranked as the highest priority ground-based astronomical facility in the 2010 NAS Decadal Survey of Astronomy \& Astrophysics \citep{NAP12951}, construction of the NSF- and DOE-funded Rubin Observatory is nearing completion. The LSST will begin deep repeated scans of the entire visible sky from Cerro Pach\'on in Chile on the same timescale (2023--2033) that many of the proposed constellation projects plan to deploy tens of thousands of LEOsats. Every night for 10 years, the LSST will take close to 1000 exposures of the deep sky with a 3200 megapixel camera (LSSTCam) covering a 9.6 square degree field of view \citep{2019ApJ...873..111I}. Because of the large collecting area, each 30 s exposure can reveal distant objects down to a limiting magnitude of 24.5 (20 million times fainter than visible with the unaided eye; \citealt{10.1093/mnras/stu992}), opening a new window on the universe.
By comparison, a typical LEOsat can be seen for several hours in twilight without the aid of a telescope, and is visible for an even longer portion of the night during summer. The rate at which a telescope-camera facility can survey the sky to a given faintness is proportional to its etendue, or the product of the telescope effective light collecting area times the angular field of view in square degrees. Rubin Observatory has the highest etendue of any existing or planned optical facility. This allows frequent repeated visits to each sky field. It is thus heavily impacted by LEOsat constellations. The number of photons collected in an exposure scales with etendue, {for both the satellite trails and all celestial objects.}  Depending on aperture and focal plane instrumentation, spectroscopic facilities may also be impacted due to their long exposures.

\subsection{How LEOsats Affect the LSST} \label{subsec:impact}

Three issues should be addressed to mitigate the effects of LEOsats on Rubin Observatory.
First, if the planned tens of thousands of LEOsats are in fact deployed, dynamic avoidance of the large number of LEOsats will be challenging. There will be some amount of lost pixel data that can be mitigated by the presence of fewer LEOsats, or by decreasing their brightness.
Second, individual LEOsats may be so bright as to affect the Rubin Observatory LSST Camera (LSSTCam) sensors, causing systematic errors in cosmological probes and resulting in fewer discoveries of near-Earth asteroids, among other scientific impacts, although these effects are not yet quantified.
Third, occasional glints of sunlight from individual LEOsats may cause a bright ``iridium flare''-like flash, which would saturate the sensors and make the entire exposure useless. For example, the SpaceX Starlink satellites exhibit these flashes at certain orientations and orbital phases, but their frequency is not yet fully known. The best mitigation option for this problem is active articulation of the spacecraft during operations.

The LSST will be different: the samples of objects will be so large that the science will be limited by systematic errors rather than sample statistics.
The science impact of LEOsat trails thus goes beyond efficiency loss (fraction of useless pixels) because key scientific investigations such as probes of the nature of dark energy and dark matter are sensitive to spatially correlated noise. Trails from bright LEOsats induce correlated noise trails at other positions on the sensor, producing a false cosmological signal. This is just one example, and we discuss several mitigation measures in this paper. A key goal is decreasing the trail brightness using direct mitigation options such as darkening or shading bright surfaces on LEOsats before launch, or to try to schedule observations to avoid the planned paths of the LEOsats.

\subsection{Rubin Observatory -- SpaceX Collaboration} \label{subsec:collab}

While these kinds of light pollution are a generic aspect of bright LEOsats, the motivation for the current study was the 2019 May launch by SpaceX of their v0.9 Starlink satellites.  SpaceX proposed to launch and operate a constellation of LEOsats at altitudes below 600 km to provide global broadband connectivity.  SpaceX currently has been granted U.S. regulatory authorization to build and operate up to 12,000 satellites, and has made international spectrum filings for an additional 30,000 satellites. This provides a unique opportunity for the current study.
In order to explore various mitigation solutions, in 2019 the Rubin Observatory Project Science Team formed a joint collaboration with SpaceX engineers working on the Starlink satellites. While it should be recognized that SpaceX is not the only source of LEOsats in operation, nor the only constellation planned, SpaceX is fielding Starlink satellites quickly, and they present the first opportunity to quantify how large numbers of LEOsats affect astronomy and learn how those effects can be mitigated by both satellite operators and ground-based observers.

\subsection{Paper Outline} \label{outline}

In this paper, we report various studies to investigate effects and mitigation strategies for Starlink satellites on the LSST. We first explore options for changing the LSST scheduling algorithm to avoid LEOsats in Section \ref{sec:leosat-survey} and simulate how LSSTCam responds to bright satellite trails in Section
\ref{sec:sim-cam-response}. We report on the SpaceX experiment to darken Starlink satellites as a mitigation strategy in Section \ref{sec:spacex-dark}. Observations of two generations of Starlink satellites are
described in Sections \ref{sec:obs-starlink} and \ref{sec:obs-darksat}. In Section \ref{sec:lab-trail-sim} we report a laboratory simulation of satellite trails and how they will impact the LSST. Finally, we summarize the status of LEOsat mitigation for the LSST in Section \ref{sec:discuss} and comment on remaining challenges in Section \ref{sec:challenges}.

\section{LEOsats and LSST Operations}
\label{sec:leosat-survey}

Most of the LSST observations will be scheduled in near-real time using a Markov decision process \citep{Naghib2019}. A robust scheduling simulation suite has been built for the LSST, incorporating a mechanical model of the telescope as well as realistic weather and downtime \citep{Delgado06}. This scheduler balances the priorities of (1) maintaining a uniform survey footprint, (2) minimizing the time spent slewing, (3) observing lower airmass regions, and (4) minimizing the number of filter changes. The scheduler is optimized using science metrics developed by the project and the general scientific community \citep{Jones2014}. An important aspect of the LSST scheduler is that outside of twilight time, the sky conditions will be relatively stable and slowly changing, allowing for $\sim40$-minute blocks of observations to be scheduled.

\begin{figure}[ht!]
\includegraphics[trim=+1cm 0 0 0, width=0.52\textwidth]{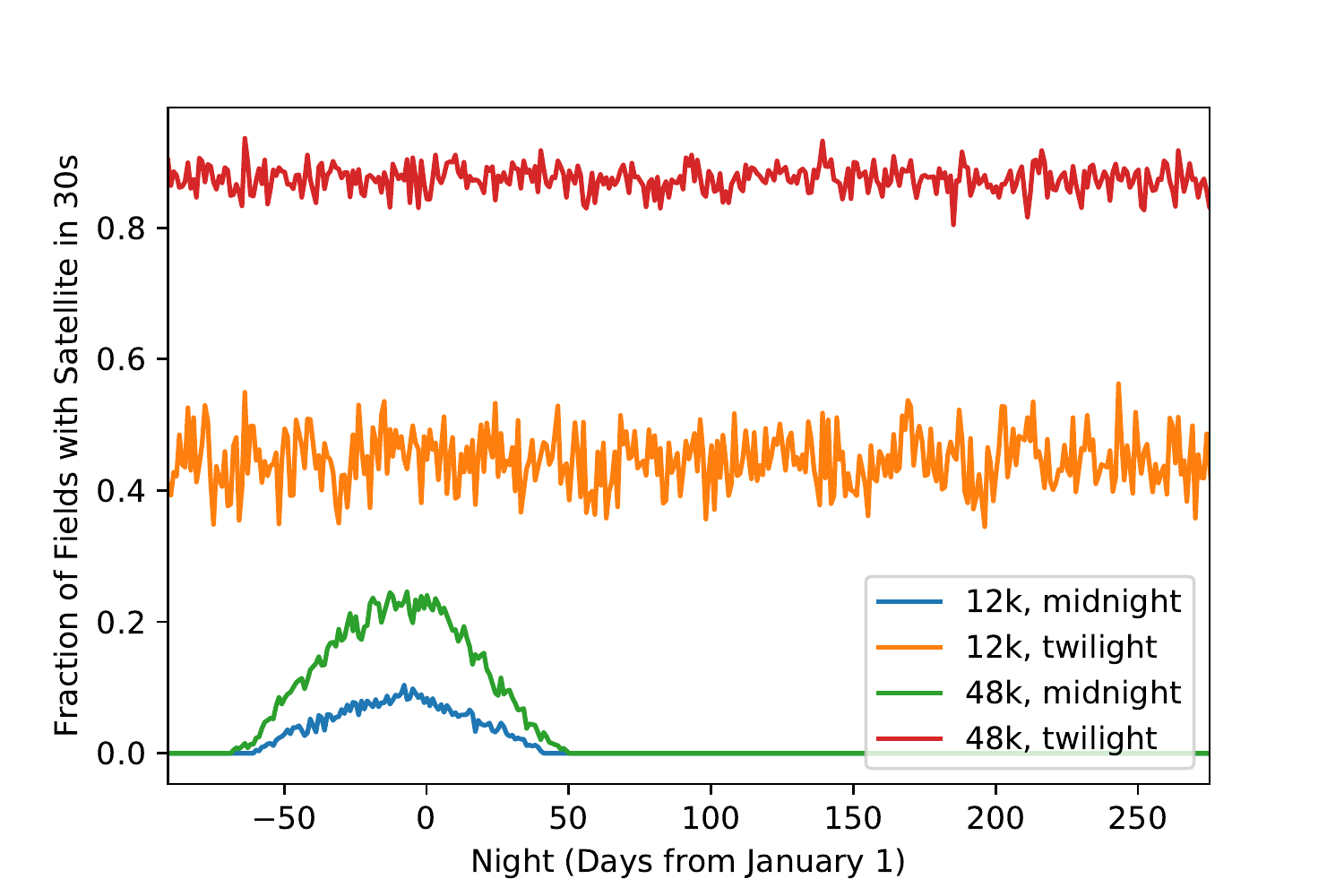}
\caption{The LSST observing scheduler was simulated for one year under two assumptions for numbers of LEOsats.
Shown here is the fraction of exposures with a satellite present versus night (number of days from January 1) at $-12$ to $-18$ deg twilight and at astronomical midnight.
Between 40\% and 90\% of exposures in normal twilight operations have an illuminated satellite trail. At midnight, the fraction of exposures with at least one satellite trail is 10--20\% during Chilean summer, and it drops to zero during Chilean winter.
\label{fig:expfraction}}
\end{figure}

Using the scheduler simulation framework, we test how a Starlink-like constellation would impact LSST observations.  We use a satellite distribution simulation developed by Benjamin Winkel\footnote{\url{https://nbviewer.jupyter.org/github/bwinkel/notebooks/blob/master/satellite_constellations.ipynb}}, and populate a range of orbital inclinations and altitudes with either the currently authorized 12,000 or the aspirational 48,000 satellites planned.
The number of illuminated LEOsats expected in LSSTCam images as a function of time of year and total constellation numbers are shown in Figure~\ref{fig:expfraction}.

At midnight in the Southern Hemisphere winter, all the Starlink satellites enter the Earth's shadow and do not leave trails. In the summer, however, a small fraction of satellites can remain illuminated even at midnight, causing 10--20\% of LSST images to have trails. With the maximum planned 48,000 satellite constellation at 550 km in place, we estimate that about 30\% of LSST images will contain at least one LEOsat trail.
\added{In this scenario, at least 0.3\% of pixels would be masked, given 0.6\% per trail per exposure, as discussed in Section~\ref{sec:sim-cam-response}. }
Typically, a satellite trail would traverse about 13--16 of the camera's 189 charge-coupled device (CCD) sensors.
In addition to 4--8 hr of nightly imaging centered on midnight, the LSST will regularly observe the sky during nautical twilight, specifically to search for near-Earth asteroids.
We find that between 40\% and 90\% of the observations taken in twilight, depending on the number of satellites, have at least one trail.
\added{The 90\% is for 48,000 LEOsats in twilight.}
\deleted{In this scenario at least 0.3\% of pixels would be masked, given 0.6\% per trail per exposure as discussed in Section~\ref{sec:sim-cam-response}. }
{For twilight observing and for 48,000 LEOsats at 300--700 km, about 1\% of the pixels would be masked.
 This is estimated via simulations of LSST twilight observing including planetary programs, using the number of trails, the angular speed of the satellite, and the width of the masks as estimated in Section~\ref{sec:sim-cam-response}.}
If in addition plans for other constellations at 1200 km materialize, they would be visible all night -- raising these numbers by an order of magnitude.  

\subsection{Satellite Avoidance Simulations} \label{subsec:dodge}
\begin{figure}[ht]
\includegraphics[trim=+0.5cm 0 0 0, width=0.46\textwidth]{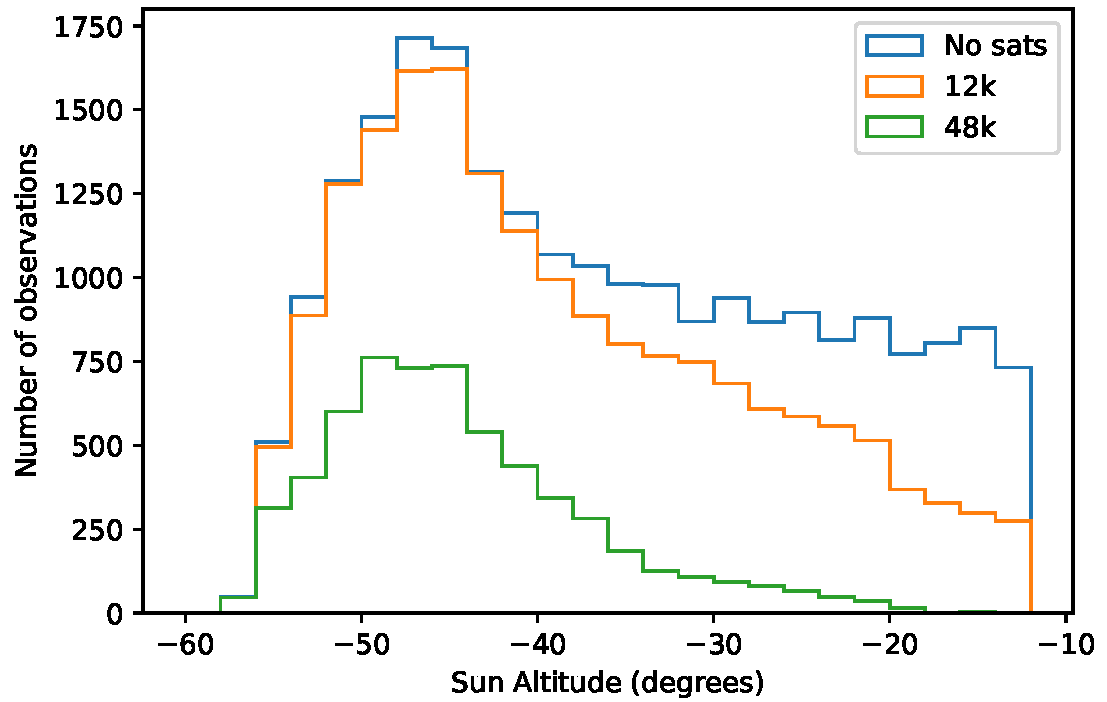}
\caption{Simulation of the number of successful observations as a function of the Sun's altitude for
30 days with active satellite avoidance. Attempts to avoid LEOsats rapidly become counterproductive as the number of LEOsats rises. For large satellite constellations, it becomes exceedingly hard to take observations that do not contain a satellite trail.
\label{fig:dodging}}
\end{figure}

We have tested a naive satellite-dodging scheme, where the observatory checks if a satellite is expected to cross during an exposure.
If a crossing will happen, the scheduler pauses for 10 s (to give the satellite time to clear the field of view) and attempts to schedule an observation again. An observation can be attempted up to three times before the scheduler abandons it and moves on to the next target in the observing queue. In the limit of very few satellites, this strategy should add a fairly negligible overhead to the night (e.g., 100 pauses of 10 sec would only result in a 3.4\% loss of efficiency). In the high satellite density limit, the scheduler will only be able to observe when it gets lucky and stumbles onto an open patch of sky. The results of this strategy are shown in Figure~\ref{fig:dodging}. The baseline survey where no dodging is attempted makes 22,662 observations over the course of 30 days.
When we attempt to avoid a 12,000 LEOsat constellation, the efficiency drops and only 18,255 observations are completed (about 80\%). For a 48,000 LEOsat constellation, only 5956 observations are completed (about 26\%). As expected, the largest hits in observing efficiency come when more satellites are illuminated. With 48,000 LEOsats, the scheduler rarely if ever finds empty areas of sky once the sun rises above an altitude of $-18^{\circ}$.

In theory, scientists could compute satellite positions ahead of time and schedule observations around them. This requires that LEOsat operators to make location data publicly available, which is not uniformly the practice in the commercial satellite industry.\footnote{We note that Starlink trajectories are presently published through \url{Space-track.org} and \url{celestrak.com}.} While this may be a useful technique for some narrow-field ground-based optical telescopes, it presents a daunting task for Rubin because of its wide field of view and because most observations need to be taken in pairs separated by $\sim20$ minutes. This is necessary to identify moving objects in the solar system, such as near-Earth asteroids. The high efficiency of the LSST scheduler comes from the ability to predominantly schedule observations of neighboring fields. For wide-field observatories like Rubin, efforts to dodge satellites while continuing to gather pairs of observations would require the scheduler to plan longer slews between observations, which is operationally inefficient. Successful preplanning of $\sim40$-minute observing blocks to avoid satellites would also require a very precise kinematic model of the telescope or require that larger inefficient overheads for slewing between pointings are included.

\begin{figure}
    \centering
    \includegraphics[width=0.48\textwidth]{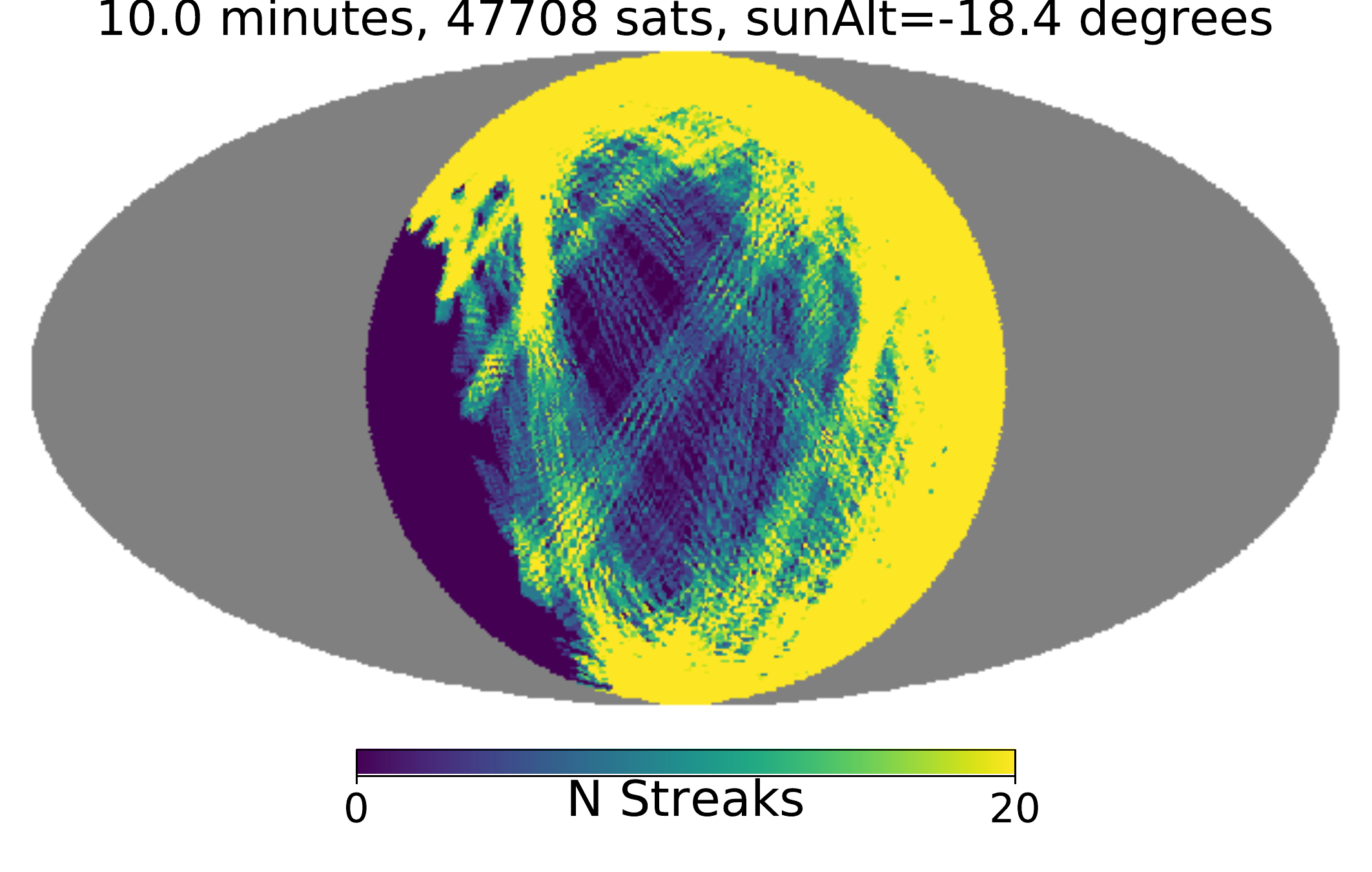}
    \caption{An all-sky Mollweide projection map showing the streaks that a mega-constellation would make over 10 minutes on a randomly chosen date (2022 October 11) just after evening twilight at the Rubin Observatory site. Zenith is at the center, north is up, and east is left. The trails are bunched because they populate the orbital planes. The trail-free region is caused by Earth’s shadow. Gray regions are below the horizon.
 }
    \label{fig:sat_steps}
\end{figure}

Figure~\ref{fig:sat_steps} illustrates the difficulty of trying to schedule observations around precomputed satellite paths. For a constellation with 48,000 satellites, over half the usually available sky area is contaminated in a 30 s exposure. In such a case, any scheduler would be forced to observe areas of the sky that are available rather than desired areas that have better conditions or have fallen behind in the survey.

\section{Simulating the LSSTCam response to satellite trails}
\label{sec:sim-cam-response}

In order to quantitatively assess the impact of satellite trails on LSST science, we must know the peak trail brightness in  $e^-$ per pixel for LEOsats as a function of satellite apparent magnitude. An LSSTCam pixel subtends $0^{\prime\prime}.2$. To address this, we computationally simulate the effect of LEOsat trails on the LSSTCam.

Illuminated by twilight, the satellite apparent brightness depends on many factors, including telescope zenith angle, distance (range), phase angle, satellite geometry, and the bidirectional reflectance distribution function (BRDF) for each component.
Our simulations are based on the latest knowledge the Rubin Observatory construction team has on the as-built system, including optical throughput of the mirrors and lenses, as well as the quantum efficiency and read noise of the detectors\footnote{\url{https://github.com/lsst-pst/syseng_throughputs}}.
The satellites are given a solar spectral energy distribution.
Our sky background model \citep{2016SPIE.9910E..1AY} is based on the ESO SkyCalc Sky Model Calculator extended to twilight using measurements from an all-sky camera on the Rubin Observatory site.

\begin{figure}[ht]
\includegraphics[trim=+1cm 0 0 0, width=0.52\textwidth]{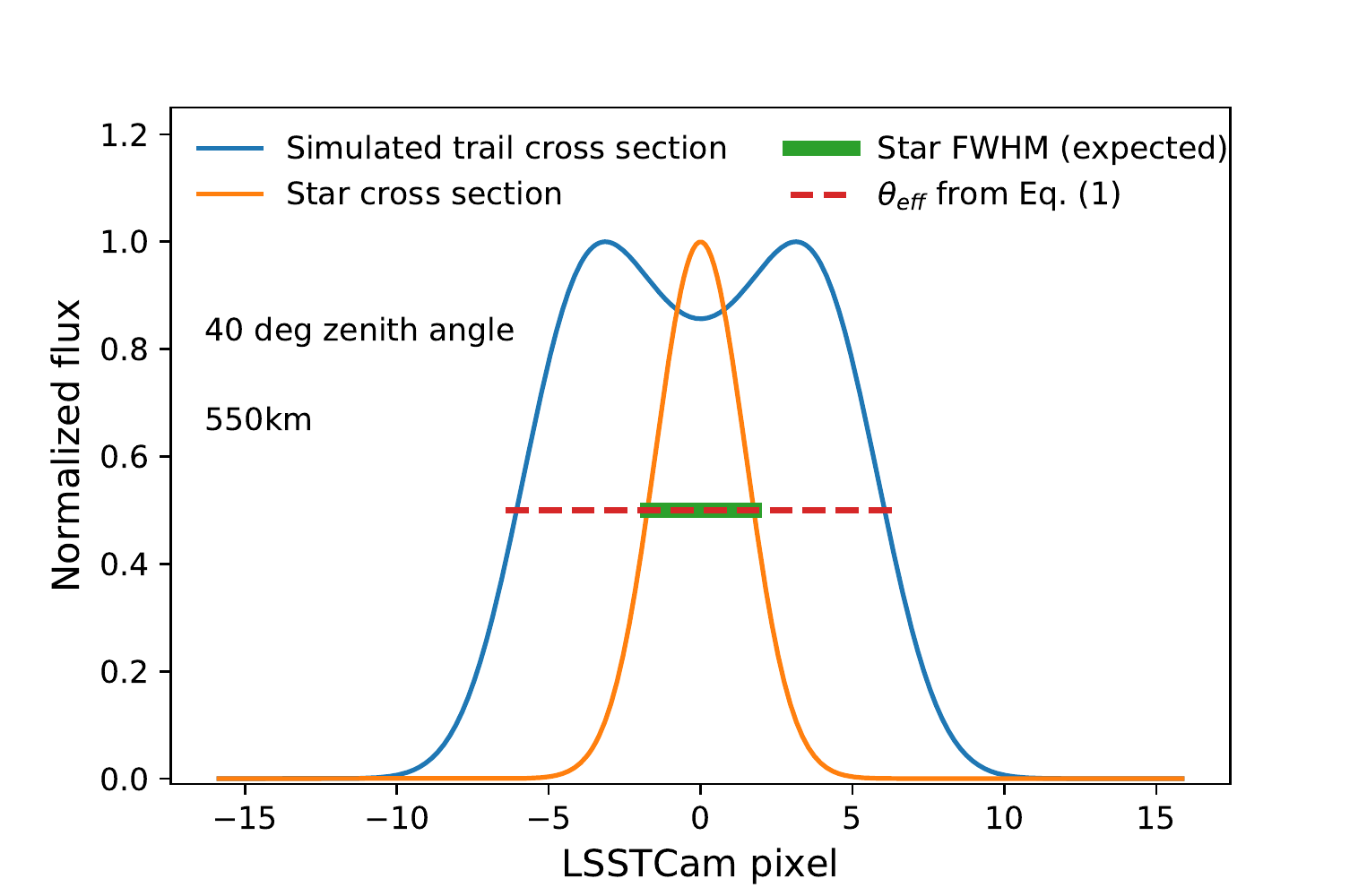}
\caption{Simulated surface brightness cross-section of a LEOsat trail at 550 km height observed at 40 deg zenith angle by Rubin Observatory. While the atmospheric seeing contributes, the dominant contribution is the angle subtended by the 8.4-m primary mirror as seen from the satellite.  \label{fig:LSSTxsection}}
\end{figure}

The surface brightness profile of a LEOsat trail $\theta_{\rm eff}$ is affected by the angular size of the satellite, the delivered seeing (typically dominated by free atmospheric seeing), and the angular size of the telescope mirror:
\begin{equation}\label{eq:satsize}
\theta^2_{\rm eff} = \theta^2_{\rm atm} + \frac{D^2_{\rm satellite} + D^2_{\rm mirror}}{d^2},
\end{equation}
where $\theta_{\rm atm}$ is the delivered seeing (in radians), $d$ is the range (distance) to the satellite, $D_{\rm satellite}$ is the satellite effective projected size, and $D_{\rm mirror}$ is the diameter of the telescope primary mirror~\citep{2018MNRAS.474.4837B}. The mirror size enters because the telescope optics are focused for parallel rays, while satellites have a finite range. A simulation of a 2 m satellite at 550 km height seen at 40 deg zenith angle with Rubin Observatory is shown in Figure~\ref{fig:LSSTxsection}.
Because of the out-of-focus effect, the instantaneous image of the satellite has a donut shape, and the transverse profile of the trail has a double-peaked structure. At 550 km height and 40 deg zenith angle, the full width at half maximum (FWHM) of the trail is about $2^{\prime\prime}.6$. For comparison, the FWHM of a typical stellar point-spread function (PSF) is about $0^{\prime\prime}.7$. Similar broad trails with square surface brightness cross section have been seen on images from the Subaru telescope~\citep{2007PASJ...59..841I}.

\begin{figure}[ht]
\includegraphics[width=0.52\textwidth]{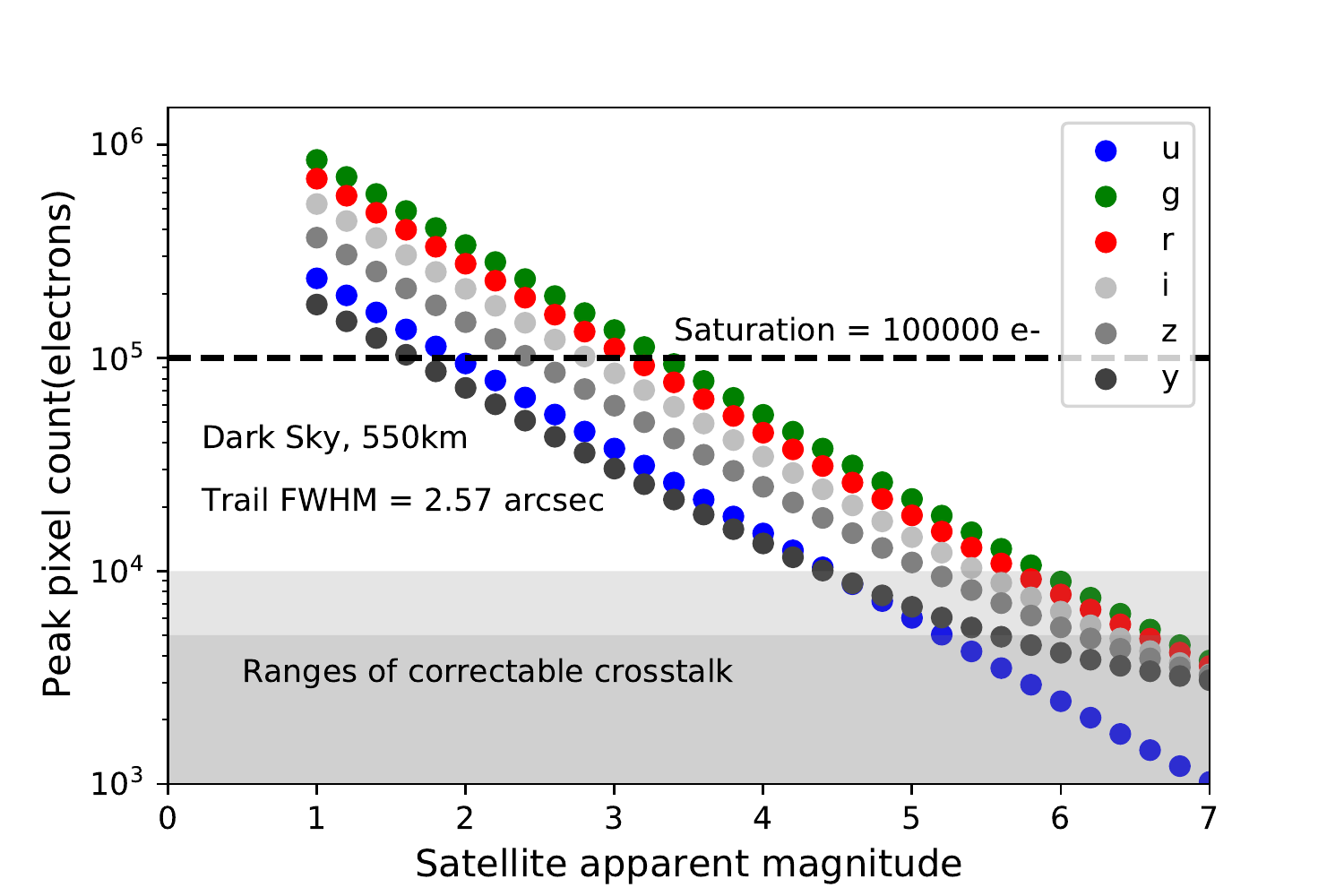}
\caption{The peak trail brightness in $e^-$ per pixel for a Starlink satellite at 550 km as a function of apparent AB mag as seen by Rubin Observatory. Colors correspond to the six different LSSTCam filter bands. The approximate saturation level of an LSSTCam CCD is indicated. The {approximate dynamic ranges over which camera crosstalk artifacts can be corrected down to below the noise level, using our current algorithm, are shown in the shaded regions} (see Section~\ref{sec:lab-trail-sim}). \label{fig:peakesim550}}
\end{figure}

In Figure~\ref{fig:peakesim550}, we show the peak counts (in $e^-$ per pixel) in the slightly resolved satellite trail versus the apparent AB magnitude in each of the Rubin Observatory optical bands for a satellite at 550 km. Saturation magnitudes vary by about 2 mag across the bandpasses ($yuzirg$, brightest in the $y$ band). The saturation level is also dependent on seeing and satellite size.
With the original brighter v0.9 Starlink satellites at $g\sim4.5$ AB mag (see Section~\ref{sec:spacex-dark}), the peak electron count is about 40,000 $e^-$ per pixel in the LSSTCam, which is then echoed across each affected CCD due to the nonlinear crosstalk of the camera readout described in  Section~\ref{sec:lab-trail-sim}.
{Figure~\ref{fig:peakesim550} shows the ranges of crosstalk correction depending on precision of crosstalk measurement for brighter satellites, as discussed in Section~\ref{sec:lab-trail-sim}. This leads to a darkening goal of 7th mag for LEOsats at 550 km.}
Various parameters from these pixel count calculations are found in Table~\ref{table:peake}.

\begin{figure}[ht!]
\includegraphics[width=0.52\textwidth]{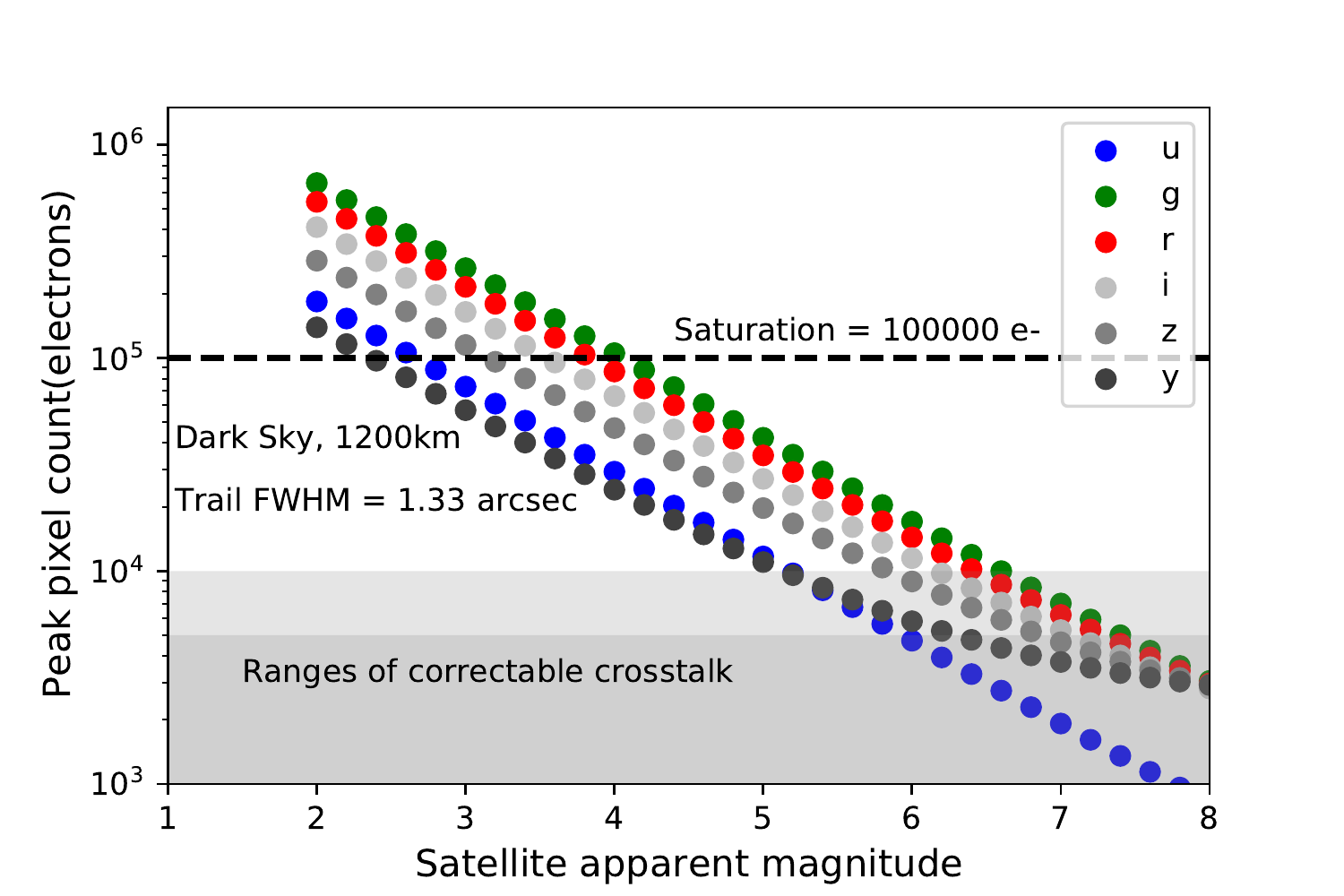}
\caption{The same as Figure~\ref{fig:peakesim550}, but for a satellite at 1200 km as seen by Rubin Observatory. Note the shift in the x-axis. For a given satellite apparent magnitude, the peak surface brightness of the trail is higher due to the smaller trail width (the satellite is more in focus). \label{fig:peaksim1200}}
\end{figure}

The large 8.4 m primary mirror helps lower the surface brightness from 100,000 $e^-$ per pixel for satellites at 550 km because they are slightly out of focus (see Equation~\ref{eq:satsize}). For comparison, we create the same plot for satellites at 1200 km in Figure~\ref{fig:peaksim1200}. At this altitude, a satellite would be more in focus.
In our simulations, we find that a LEOsat at 1200 km would have to be $g\sim8$ mag or fainter in order to be well within the range of correctable crosstalk because the trail is less spread out.

\begin{deluxetable}{lrrrrrr}
\centering
\caption{Parameters from Peak Pixel Count Calculations. \label{table:peake}}
\tablehead{
 \colhead{} &
 \colhead{$u$} &
 \colhead{$g$} &
 \colhead{$r$} &
 \colhead{$i$} &
 \colhead{$z$} &
 \colhead{$y$}
 }
\startdata
    $m_{\rm sky}$$^a$             &      22.96  & 22.26  & 21.20  & 20.48  & 19.60  & 18.61  \\
    $N_{\rm sky}$$^b$            & 81  & 411  & 819  & 1173  & 1783  & 2371    \\
    $T_b$$^c$   & 0.036  & 0.129  & 0.105  & 0.080  & 0.055  & 0.027   \\
     $m_{\rm sta}$$^d$     & 1.50  & 2.89  & 2.67  & 2.37  & 1.98  & 1.19     \\
    $m_{\rm tra}$$^e$   & 14.27  & 15.66  & 15.44  & 15.15  & 14.75  & 13.96     \\
    $m_{\rm sta}^{\rm X}$$^f$ & 4.44  & 5.87  & 5.70  & 5.44  & 5.12  & 4.41  \\
\enddata
\tablecomments{The satellite is assumed to be at 550 km height, $40^{\circ}$ zenith angle, with an apparent size of 2 m and angular speed of 0.5 deg sec$^{-1}$. The exposure time is 30 s. Since the trail width for LSSTCam is dominated by the primary mirror size, we use $0^{\prime\prime}.7$ seeing in all the bands. \\
$^a$ Expected sky brightness at Cerro Pach\'on (AB mag arcsec$^{-2}$) based on \cite{2016SPIE.9910E..1AY}.  \\
$^b$ Sky counts ($e^-$ per pixel) corresponding to $m_{\rm sky}$ and 30 sec exposure. \\
$^c$ Throughput integral, $T_b = \int S^{\rm atm}(\lambda) S^{\rm sys}_b(\lambda) \lambda^{-1}d\lambda$, where $\lambda$ is the wavelength, $S^{\rm atm}(\lambda)$ is the atmospheric throughput, and $S^{\rm sys}_b(\lambda)$ is the system throughput in each band. \\
$^d$ Satellite stationary magnitude whose peak pixel count reaches the saturation level of 100,000 $e^-$. \\
$^e$  Satellite trail surface brightness (AB mag arcsec$^{-2}$) corresponding to $m_{\rm sta}$.\\
$^f$  Satellite stationary magnitude whose peak pixel count reaches the approximate \added{best-case} crosstalk correctable limit of 10,000 $e^-$.}
\end{deluxetable}


For given LEOsat stationary magnitudes and exposure time, we can predict the DECam ADU counts and the trail surface brightness.
As a constency check, we carry out the same simulations for the Dark Energy Camera (DECam) using public filter throughput data\footnote{\url{http://www.ctio.noao.edu/noao/content/decam-filter-information}}.  These results agree well with the real measurements, which are presented in Section~\ref{sec:obs-darksat}.

\added{Our current baseline approach is to mask the trails in the data products. As discussed in Section~\ref{sec:challenges}, the residual surface brightness systematic error target is 1 $e^-$ per pixel for correlated pixels along a line. While a stack of $\sim$100 images in a band will be input to the detection and photometry, the actual masking is done on individual single exposures. Variants of the Hough transform have been used to detect and mask trails~\citep{ppHough, modHough, DESpipe}. Masking algorithms that automatically mask pixels along a trail above $5 \sigma$ of sky noise create parallel lines of correlated pixels at the mask edge, which are then diluted in the coadd. The residual correlations in these linear features are near the outer surface brightness of the faint galaxies used for cosmic shear. Growing the mask to suppress this to 10\% of the surface brightness of these faint galaxies leads to a conservative mask threshold of 1\% of the sky noise.}

\begin{figure}[ht]
\includegraphics[width=0.52\textwidth]{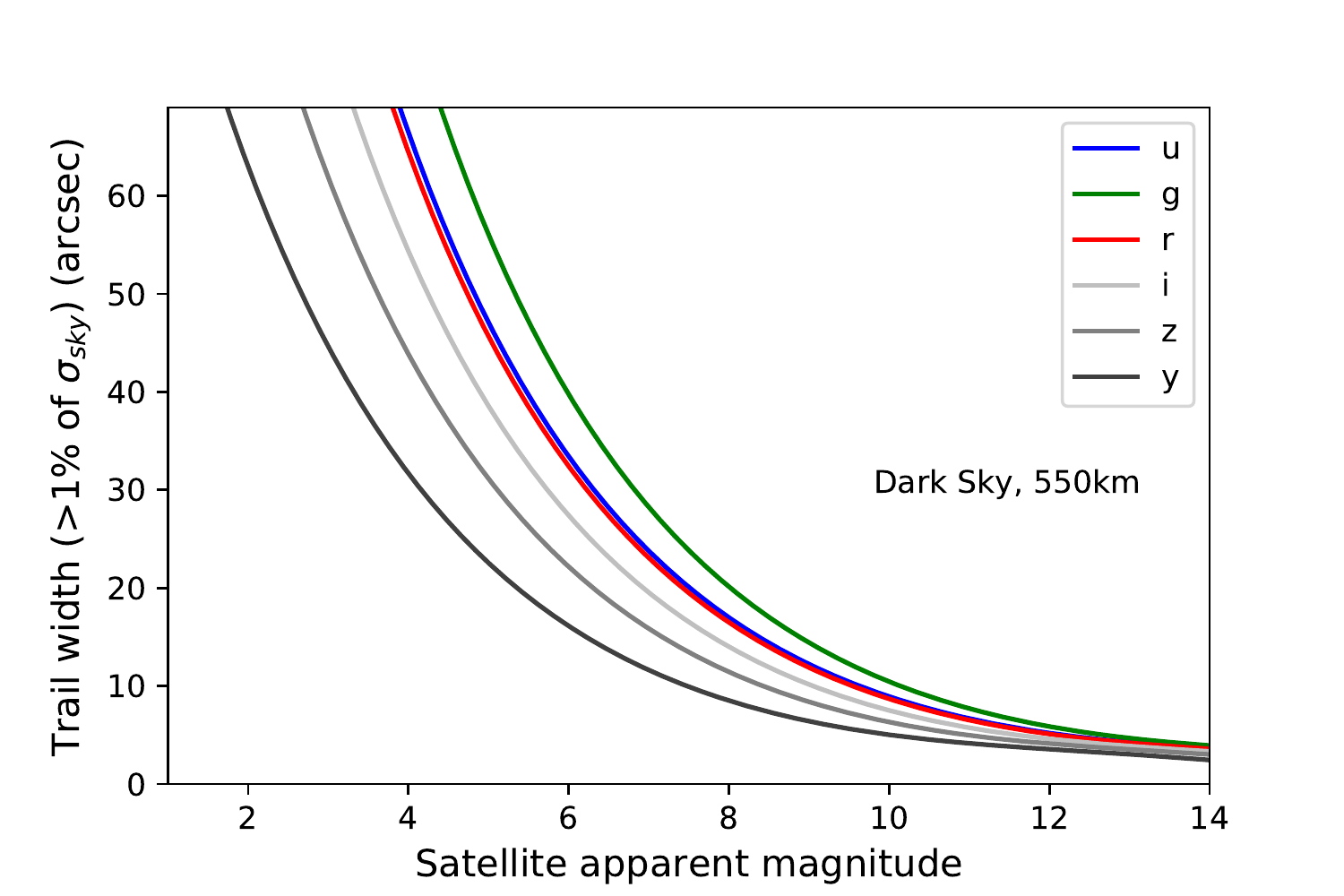}
\caption{The width of a satellite trail over which the streak counts exceed {1\%} of the sky background noise ($\sigma_{\rm sky}$) as a function of apparent AB mag as seen by the Rubin Observatory.
Conservative masking to 1\% of sky noise results in trail widths of $60^{\prime\prime}$ in $g$ band for 4.5 mag satellites.
\label{fig:widthsim550}}
\end{figure}

The width of the satellite trail at a given surface brightness depends on the PSF, the satellite apparent magnitude, its size and range, and the size of the telescope mirror. We show in Figure~\ref{fig:widthsim550} the width of a trail from a 2 m satellite at 550 km and range 1000 km at a surface brightness corresponding to {1\%} of the sky background noise for a 30 s exposure in six bands for the Rubin Observatory.  The seeing profile is simulated with a von K\'{a}rm\'{a}n turbulence model~\citep{vk1, vk2} with an outer scale of 30 m.
The uncertainty of the sky background is taken as $\sigma_{\rm sky}=\sqrt{N_{\rm sky}}$, where $N_{\rm sky}$ is the sky surface brightness expressed in counts per pixel; the values used in our calculations are listed in Table~\ref{table:peake}.
At the apparent magnitude of current LEO satellites, approximately 0.6\% of the pixels in the LSSTCam  would have to be masked per exposure per trail (we have used a trail width of $1^{\prime}$ in this calculation, which corresponds to a 1\% level of the sky noise).
In a 30 s visit for the LSST, a typical satellite would have traveled $\sim$15 deg, which is much larger than the 3.5 deg diameter of the field of view (FOV).  For an image with a single trail, the fraction of lost pixels is proportional to the ratio of the trail width to the FOV diameter. Therefore the fraction of lost pixels in all LSST images is independent of etendue and increases linearly with exposure time and the trail width. Many  exposures during twilight will have multiple trails (the expectation value for LSSTCam is about two trails per twilight exposure). The science impact of the LEO satellites, which is of course also proportional to the total number of satellites, goes much beyond the fraction of lost pixels.

\section{SpaceX experiments darkening Starlink satellites}
\label{sec:spacex-dark}

Prior to the current generation of LEOsat constellation deployments, the periodic and short-duration visibility of LEOsats in the hours shortly after nautical twilight was considered to be more of a curiosity, and only a minor nuisance to astronomical observation.
The SpaceX 2019 May launch of 60 v0.9 Starlink satellites in a single deployment yielded a level of visibility and impact on optical observations that surprised both the astronomical community and designers of satellite constellations. Trains of the 60 starlink satellites were visible to the unaided eye, and appeared as parallel trails in sidereal-tracking, wide-field, long-exposure ground-based astronomical observations\footnote{\url{https://nationalastro.org/news/starlink-satellites-imaged-from-ctio}}. Because the Starlink satellites do not emit visible light, sunlight reflected from the satellite is the source of the observed visible signature.

Solar radiation is a double-edged sword from the standpoint of satellite design. While the solar array generates all spacecraft power from sunlight, solar radiation presents a significant thermal load for nonarray satellite components. This load is typically reduced by decreasing the absorptivity of external surfaces. Solar radiation brightness peaks at $\sim$555 nm, or the center of the visible band. Reduced absorptivity in the visible band results in an increased optical signature of the satellite because conservation of energy requires nonabsorbed light to reflect.  Ignoring the effects of thermal transients and close coupling to Earth, the equilibrium temperature of a notional, spherical, sunlit graybody, involves a balance between between absorbed sunlight, electronics heating, and thermal radiation ($\sim9-10~\mu$m) to deep space. The equilibrium temperature $T$ of the satellite may be written as
\begin{equation}
T = \left[\frac{\alpha S}{4\epsilon_{\rm IR} \sigma}(1+f)\right]^\frac{1}{4},
\end{equation}\label{starlink-temp}
\noindent where $\alpha$ and $\epsilon_{\rm IR}$ are the values of emissivity $\epsilon(\lambda)$, weighted using incoming solar flux and appropriate thermal infrared emission flux, respectively, where $S$ is the solar flux $(\sim1360 \textrm{W m}^{-2})$, $\sigma$ is the Stefan-Boltzman constant, and $f$ is the ratio of the satellite component power dissipation (primarily electronics) to the absorbed power from the projected area of sunlight illuminating the bus. Reducing the $\alpha S$ product reduces the equilibrium temperature, but results in an increased optical signature (more reflected solar light).

A key tool for satellite thermal control is the radiator, which has the dual purpose of reducing solar absorption (i.e., reflecting) sunlight while maintaining a high emissivity in the thermal infrared band. Reflected light may be broken down into two basic classes: specular (mirror-like reflection) and diffuse (reflections spread over a wide solid angle). A specular reflection is observed in only one direction, while a diffuse reflection may be observed in any direction from the surface (albeit at a much lower intensity than a specular reflection). A specular or diffuse surface can be generally categorized as white ($>$90\% reflectivity) or black ($<$10\% reflectivity).

The original v0.9 Starlink satellites had diffuse white external surfaces comprised of bare metal, anodized aluminum, white electronic components, and dedicated radiator surfaces (composed of an optically transparent outer layer that radiates in the thermal infrared and an optically reflective inner layer to reject sunlight). Initially, the optically reflective thermal radiator surfaces were thought to be the main source of reflected sunlight, but observations of the satellite under directional light showed the radiator surfaces to be dark, with the majority of visible light reflecting from the diffuse white surfaces (antennas and bare metal).

\begin{figure}[ht]
\includegraphics[trim=+1cm 0 0 0, width=0.47\textwidth]{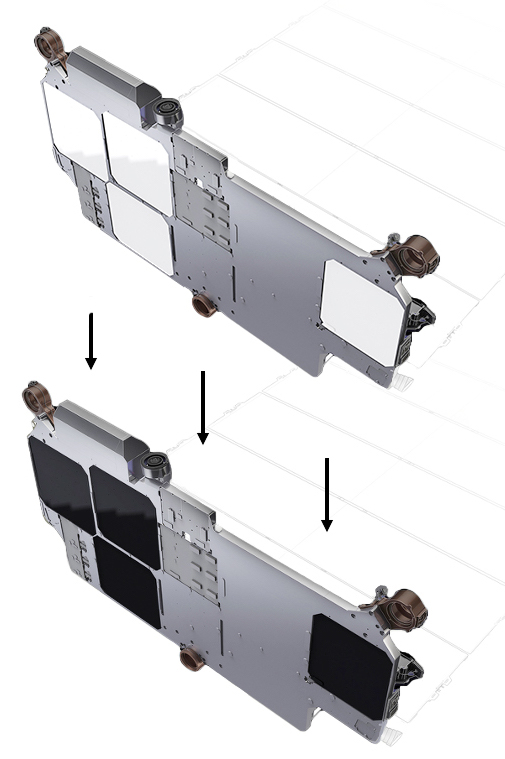}
\caption{A cartoon showing the four phased arrays on DarkSat which were darkened to reduce diffuse reflection. This was in addition to other darkening measures discussed in the text.\label{fig:darksat}}
\end{figure}

An experimental satellite (Starlink-1130, or ``DarkSat'') was launched in 2020 January. Previously white satellite surfaces were covered with either black diffuse applique or the thermal radiator surfaces mentioned previously. Communication elements (antennas) were painted with a specular black paint. Additionally, 18 of the other 59 satellites launched with Starlink-1130 had previously bare metal elements covered with the thermal radiator material. Four of the darkened phased array antenna panels are shown in Figure~\ref{fig:darksat}.

An alternative method of darkening the white diffuse phased array panels is to use an external, radio-frequency transparent ``sunshade'' to block sunlight from reaching the white panels. This has the added benefit of reducing the equilibrium temperature of the phased array antennas. An experimental follow-on satellite (``VisorSat'') will be the subject of future observations and analysis.

To understand the origin of the remaining sources of optical brightness of the satellite, SpaceX is developing an optical signature model that moves beyond simple Lambertian scattering and phase approximations.  The model combines CAD geometry and source material BRDF measurements to predict the radiant intensity profile of each component, calibrated via material sample measurement and ground-based observations. This will be a useful tool in predicting the apparent brightness of the complex satellite shape as a function of solar illumination and observer location.

The multiple changes to the spacecraft discussed here, plus new mitigations, will be implemented in future Starlinks.

\section{Observations of a v0.9 Starlink}
\label{sec:obs-starlink}

In late 2019 May, SpaceX launched the first 60 of its planned constellation of 12,000 LEOsats. For optical astronomy, the noted concerns are the number and brightness of satellite trails, and the anticipated effects on survey data.
In order to assess the LEOsat brightness impact on the Rubin Observatory, Todd Boroson of Las Cumbres Observatory (LCO) Global Observatory\footnote{\url{https://lco.global}} obtained repeated photometry on one v0.9 Starlink satellite at its 550 km operational altitude (private communication).

Using several LCO Global 1 m and 40 cm telescopes instrumented with $20^{\prime} \times 30^{\prime}$ field CCD cameras, Boroson made eight attempts to observe Starlink-81 (NORAD 44292), all in the $V$ band, from 23 to 72 deg altitude. A trail from the satellite was detected in four of the images in 3 arcsec seeing. These showed integrated $V$ apparent magnitudes from 5.8 to 7.6, all between 70 and 95 minutes from sunrise or sunset. Two attempted observations of the satellite about 3.5 hr before sunrise did not detect it. It is important to point out that the peak surface brightness of a satellite trail (above the sky level, measured in units of  $e^-$ per pixel) is independent of the exposure time as long as the exposure time is longer than the time it takes the satellite to trail across the field of view.

To calibrate ADU per sec per $V$ mag, Boroson measured the total flux in the trail over many pixels, and then divided by the time that it took the satellite to travel over that many pixels. The resulting surface brightness is the equivalent magnitude over $1^{\prime\prime}$ of the trail. All the measurements were calibrated via stars of known brightness in the fields.

Extrapolated to zenith, this v0.9 Starlink was 4.5--5 $g$ AB mag. We can extrapolate this calibrated photometry to the peak trail surface brightness that the Rubin Observatory LSSTCam would see. After correcting for the better $0^{\prime\prime}.7$ seeing on Cerro Pach\'on and for the larger telescope primary mirror, we found that this satellite at zenith would appear sufficiently bright to generate artifacts in LSSTCam images, above 50,000 e$^{-}$ pixel$^{-1}$. These initial observations informed and motivated the laboratory measurements on LSSTCam CCDs described in Section~\ref{sec:lab-trail-sim}.

\section{Observations of DarkSat}
\label{sec:obs-darksat}

Recently, \citet{2020A&A...637L...1T} reported photometry of Starlink DarkSat in the $r$ band with a 0.6 m telescope. They find that when scaled to a range of 550 km and corrected for the solar and observer phase angles, a reduction by a factor of two is seen in the reflected solar flux between DarkSat and one of its siblings on the same launch, Starlink-1113.

We report here an analysis of Starlink $g$-band observations obtained using the Blanco 4 m telescope DECam resulting from our Director's Discretionary time application, which obtained data during observations for the DELVE Survey\footnote{\url{https://delve-survey.github.io}} led by Alex Drlica-Wagner. Five Starlink satellites are studied. As described below, we find that DarkSat is 1.1 mag fainter than its closest companion.

\subsection{Photometry of five recent Starlink satellites}
\label{subsec:decam-photometry}

The Blanco 4m telescope was pointed at coordinates provided by SpaceX that corresponded to the predicted peak altitude of each satellite's path across the sky. The shutter was opened approximately 30 s before the time of that prediction, and a 120 s exposure combined with the DECam one-degree-wide field of view guaranteed capture of the satellites' passage. Thin clouds were present throughout, and seeing was approximately $1^{\prime\prime}.2$. Trails of five satellites were acquired in four $g$-band images taken during twilight hours around midnight UTC on 2020 March 6 (local time 21:05--21:35 on 2020 March 5, about one hour after sunset). One visit includes trails from both DarkSat and one of its brighter siblings (Starlink-1112).

Raw visit images, bias frames, and dome flats were retrieved from the NOAO Science Archive\footnote{\url{http://archive1.dm.noao.edu}}. We used the LSST Science Pipelines \citep{2019ASPC..523..521B} to build a master bias and master flat and perform instrument signature removal. The image background and typical PSF were then modeled, the background was subtracted, and the images were astrometrically and photometrically calibrated using reference catalogs from Gaia DR2 \citep{2018A&A...616A...1G} and Pan-STARRS1 \citep{2016arXiv161205243F}, respectively. Each visit consists of 60 CCDs with usable image data. We selected one CCD for each satellite---the one with the longest trail---for further analysis. The (visit, CCD) pairs used are (941420, 7), (941422, 33), (941424, 34), (941424, 37), and (941426, 16). These correspond to Starlink-1102, -1073, -1130 (DarkSat), -1112, and -1084, respectively.

To measure properties of each satellite trail, we manually identify two points at opposite ends of the trail in a single CCD image. The image is then rotated to make the trail appear horizontal, and we analyze a horizontal stripe 40 pixels wide centered on the brightest part of the trail. Because the images have been photometrically calibrated, the pixel values in this stripe are in units of nJy. We use this together with the DECam pixel scale to sum the pixel values and report the raw trail brightness in mag arcsec$^{-2}$. We then compute a ``corrected'' trail brightness to account for the 120 s exposure time.

We compute each satellite's angular speed in the sky assuming a height of 550 km using the airmass, orbital speed from a circular orbit, and the angle between the trail and the horizon. These speeds are all between 0.5 and 0.8 deg s$^{-1}$, and we verify that they agree with estimates from sparse SpaceX telemetry to within 10\%. We combine this with the DECam pixel scale and the average FWHM of a set of Gaussians fit to each pixel slice of the trail profile to compute a stationary satellite magnitude.

Next, we extrapolate how bright the satellite would be if it appeared at zenith by subtracting $5 \log_{10} (\textrm{airmass})$, which accounts for flux variation with distance. Finally, we report the derived satellite size $D_{\rm satellite}$. The size is computed from Equation~\ref{eq:satsize} by subtracting the FWHM for the PSF from the trail's FWHM in quadrature, multiplying by the airmass to extrapolate to zenith, and subtracting the contribution from the size of the telescope mirror. This analysis is publicly available on GitHub\footnote{\url{https://github.com/dirac-institute/starlink}}. Table~\ref{tab:delve} summarizes these measurements. All magnitudes are in the AB system.

\begin{deluxetable*}{lhhccccccccccccc}
\tabletypesize{\scriptsize}
\tablecaption{Five Starlink Satellites Imaged in $g$ Band with DECam on the Blanco 4 m in 2020 March\label{tab:delve}}
\tablewidth{0pt}
\tablehead{
\colhead{Starlink} &
\nocolhead{Visit} &
\nocolhead{CCD} &
\colhead{Time} &
\colhead{Phase} &
\colhead{Airmass} &
\colhead{PSF} &
\colhead{Background} &
\colhead{Trail} &
\colhead{Raw trail} &
\colhead{Corrected trail} &
\colhead{Speed} &
\colhead{Stationary} &
\colhead{Zenith} &
\colhead{$d$} &
\colhead{Size} \\
\colhead{} &
\nocolhead{} &
\nocolhead{} &
\colhead{} &
\colhead{Angle} &
\colhead{} &
\colhead{FWHM} &
\colhead{} &
\colhead{FWHM} &
\colhead{(mag} &
\colhead{(mag} &
\colhead{} &
\colhead{} &
\colhead{} &
\colhead{} &
\colhead{} \\
\colhead{} &
\nocolhead{} &
\nocolhead{} &
\colhead{(UTC)} &
\colhead{(deg)} &
\colhead{} &
\colhead{(arcsec)} &
\colhead{(mag arcsec$^{-2}$)} &
\colhead{(arcsec)} &
\colhead{arcsec$^{-2}$)} &
\colhead{arcsec$^{-2}$)} &
\colhead{(deg s$^{-1}$)} &
\colhead{(mag)} &
\colhead{(mag)} &
\colhead{(km)} &
\colhead{(m)}
}
\startdata
1102 & 941420 &  7  & 00:05 & 56.2 & 1.03 & 1.35 & 19.0 & 2.43 & 19.98 & 14.78 & 0.77 & 5.21 & 5.15 & 565 & 3.84  \\
1073 & 941422 & 33  & 00:15 & 56.4 & 1.15 & 1.35 & 19.2 & 2.04 & 19.96 & 14.76 & 0.70 & 5.49 & 5.18 & 625 & 2.34  \\
1130 & 941424 & 34  & 00:30 & 60.1 & 1.55 & 1.20 & 18.9 & 2.12 & 21.31 & 16.11 & 0.54 & 7.08 & 6.13 & 810 & 5.58  \\
1112 & 941424 & 37  & 00:30 & 60.1 & 1.55 & 1.18 & 19.0 & 1.87 & 20.06 & 14.86 & 0.54 & 5.97 & 5.02 & 810 & 4.02  \\
1084 & 941426 & 16  & 00:35 & 61.2 & 1.71 & 1.33 & 18.8 & 1.82 & 20.27 & 15.07 & 0.50 & 6.29 & 5.13 & 878 & 3.47  \\
\enddata
\tablecomments{All exposures are from 2020 March 6 UTC with a 120 s exposure time. The distance to the satellite $d$ and derived satellite size ``Size'' correspond to $d$ and $D_{\rm satellite}$ from Equation~\ref{eq:satsize}, respectively. Starlink-1130 is DarkSat.}
\end{deluxetable*}

\subsection{DarkSat compared to its brighter siblings}
\label{subsec:darksat-compare}

A key value in Table~\ref{tab:delve} is the column ``Stationary mag,'' which is the magnitude the satellite would have if it were not moving. The following column, ``Zenith mag,'' is the stationary magnitude extrapolated to zenith.

\begin{figure}[ht]
\includegraphics[trim=+1cm 0 0 0, width=0.52\textwidth]{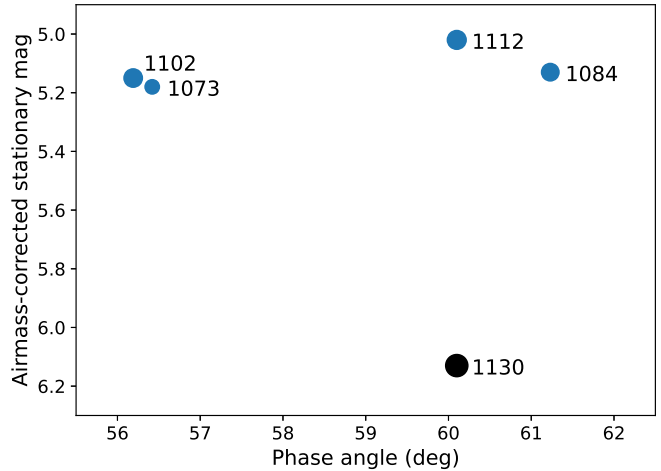}
\caption{Apparent stationary $g$ magnitude of five recent Starlink satellites extrapolated to zenith as a function of solar phase angle. DarkSat (black) is 1 mag fainter than its four bright siblings (blue), which are in turn about 0.5 mag fainter than the v0.9 Starlinks. The point sizes correspond to each satellite's derived size from Table~\ref{tab:delve}. \label{fig:5sats}}
\end{figure}

Figure~\ref{fig:5sats} illustrates the reduction of brightness by 1.1 mag for DarkSat (Starlink-1130) compared to Starlink-1112 which was observed in the same visit. The satellites that we call ``bright siblings'' here are fainter by about 0.5 mag than the original v0.9 Starlink satellites (Boroson's measurements at LCOGT).
The difference is due to a change from diffuse reflection (by aluminum surfaces) to specular reflection due to mitigation efforts described in Section \ref{sec:spacex-dark}. Each of the four siblings of Starlink-1130 had previously bare metal elements covered with the thermal radiator material. This explains why they are nearly equal in apparent magnitude when normalized to zenith. To within error, these measurements are consistent with those of \citet{2020A&A...637L...1T}.

All satellite trails in these data analyzed here are widened by the effects described in Section~\ref{sec:sim-cam-response} and shown in Equation~\ref{eq:satsize}. The observed trail FWHM is the sum in quadrature of the seeing (PSF FWHM in Table~\ref{tab:delve}), the angular size of the satellite, and the angular size of the telescope mirror. The last term is due to the telescope being focused for parallel rays coming from a source at infinity. The derived satellite sizes at zenith (Size in Table~\ref{tab:delve}, or $ D_{\rm satellite}$) corresponds to $\sim3$ m at 550 km, which agrees with expectations.

\begin{figure}[ht]
\includegraphics[trim=+1cm 0 0 0, width=0.52\textwidth]{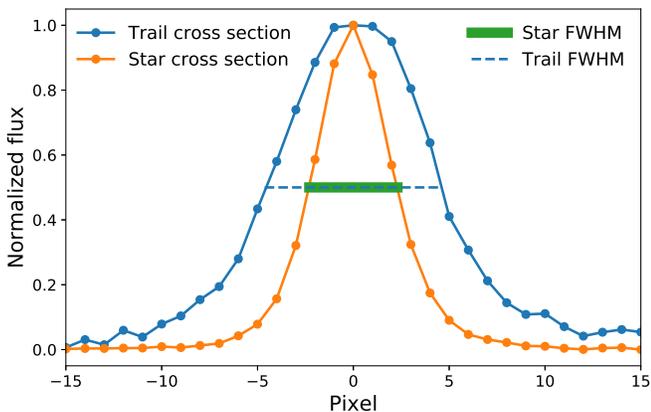}
\caption{Apparent surface brightness cross section of the Starlink-1102 trail as observed by DECam on the Blanco 4-m telescope, in blue. This visit had an airmass of 1.03 (zenith angle $\sim14$ deg). The star cross section shown in orange is the PSF kernel derived from fitting all the identified stars in the image. While the delivered seeing (PSF) contributes to the satellite profile, the dominant contribution is the angle subtended by the 4 m mirror as seen from the satellite. This effect will be even larger with the Rubin Observatory 8.4 m mirror, resulting in a wider trail as in Figure~\ref{fig:LSSTxsection}. \label{fig:xsection}}
\end{figure}

Figure~\ref{fig:xsection} shows the surface brightness profile of Starlink-1102 in the 4 m Blanco telescope data as well as a typical stellar PSF profile for reference. Both are normalized to unity. From Table~\ref{tab:delve}, this observation was taken when the satellite was about 14 deg from zenith and the solar phase angle was 56 deg. Assuming a 550 km orbit altitude, the distance $d$ from the telescope to the satellite was 565 km. Figure~\ref{fig:xsection} is the real Blanco telescope equivalent of Figure~\ref{fig:LSSTxsection}, which shows the same profile comparison as simulated for Rubin Observatory. In both cases, the satellite trail is wider than the PSF profile.

\section{Laboratory simulations of bright satellite trails on LSSTCam CCDs}
\label{sec:lab-trail-sim}

To better understand systematic effects of bright linear features on LSSTCam images, we began laboratory tests on science-grade CCDs in early summer 2019. While we carried out tests using two separate systems with differing readout electronics, we describe results from a LSSTCam CCD hardware beam simulator here. The first laboratory beam simulator imaging campaign preceded the LSSTCam testing, and we obtained similar initial results on multiple LSSTCam CCDs.

Various methods for simulating the trail of a satellite have been tested, including diagonal and linear dithering of bright spots and lasers, as well as projector systems and photolithographic masks. So far, the most realistic satellite streaks have come from using the LSST $f/1.2$ re-imaging facility \citep{2014SPIE.9154E..15T} to re-image a $\sim40~\mu$m wide slit on a science-grade LSST e2v CCD, where we use an optical beam identical to LSSTCam to form a line about four pixels wide extending across most of the detector. This does not use the LSSTCam electronics, but the crosstalk effects seen are very similar to those with the LSSTCam tests. We took several thousand exposures at various illumination levels going from 100 $e^-$ to 250,000 $e^-$ per pixel, along with random slit mask rotations in order to collect data representing random LEOsat crossings across many revisits to a field.

The result of our tests unsurprisingly indicate that LEOsat trails cause many undesirable image artifacts in the CCD data. The severity of the artifacts depends on the brightness of the satellite compared to CCD saturation. Earlier simulations of the LSSTCam optics showed that at very bright levels, corresponding perhaps to zeroth-magnitude flashes or glints of sunlight off spacecraft surfaces, the satellite can cause scattered light within the telescope optics and the cryostat, and blooming of charge across the CCD. Entire exposures, or at least large segments of the focal plane, would be lost. However, this should be an extremely rare ($10^{-4}$ per satellite pass, although this is not well known; \citealt{2020A&A...636A.121H}) occurrence for LEOsats only at certain orientations and orbital phases. We anticipate that the net impact on LSST would be negligible.

With satellite trails below CCD saturation, the main concern is crosstalk of the trail into neighboring channels of a CCD. Each CCD has 16 one-megapixel segments that are each read out by independent, parallel processing channels. These ``video'' channels traverse cables in close physical proximity, and are processed simultaneously at 500,000 pixels s$^{-1}$, causing low-level coupling between the channels within a CCD. It is also possible that some crosstalk originates in the readout electronics, which amplifies the video signals and executes correlated double sampling with dual slope integration.
As mentioned above, the tests we carried out involve different electronics. The LSSTCam system incorporates an application-specific integrated circuit (ASIC), and while tests of that system have only begun, some contribution to the nonlinear crosstalk appears to originate in the ASIC.

For both systems, this crosstalk coupling appears to be a hundredth of a percent at worst ($10^{-4}$) and $<10^{-6}$ at best between well-separated channels. In contrast to classic capacitive coupling, it can also be negative and nonlinear with respect to the main trail signal. This nonlinear behavior with flux is new and noteworthy. This unavoidable crosstalk means that trails left by satellites have a multiplicative effect, causing the appearance of lower level ``electronic ghosts'' alongside the main trail.

\begin{figure}[ht!]
\centering
\includegraphics[width=0.4\textwidth]{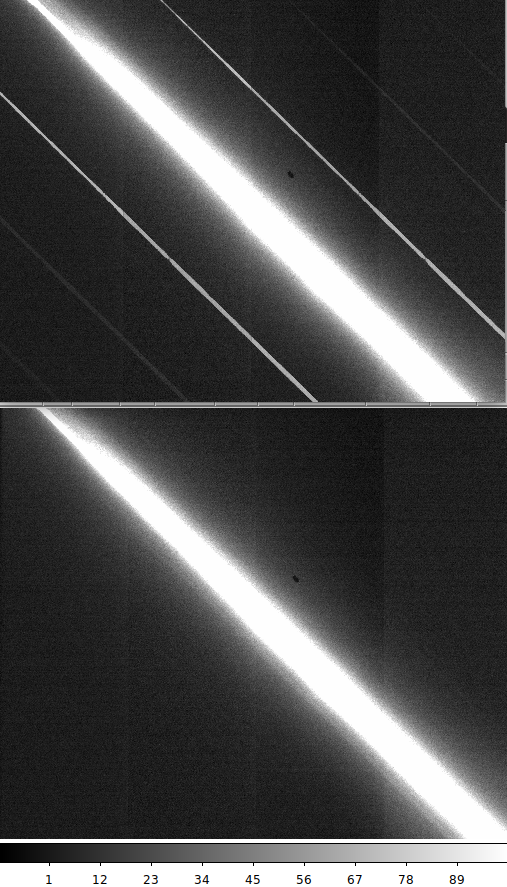}
\caption{Top: the image that results when an artificial satellite trail at the level corresponding to v0.9 Starlink satellites (bright, but below pixel saturation) is projected onto a e2v CCD in the laboratory. Four of 16 channels of a single raw CCD image are shown, and six crosstalk stripes induced by the main trail are visible. Below: the same image after a preliminary nonlinear crosstalk correction algorithm has been applied (see Section~\ref{subsec:crosstalk}). While the crosstalk trails are nearly removed, the remaining trail itself is several hundred pixels wide and has a surface brightness $\sim$1000 times that of important astrophysical signals. \label{fig:xtalk}}
\end{figure}

The top panel of Figure~\ref{fig:xtalk} shows a 4-megapixel cutout of an LSST e2v 16-megapixel CCD image of with a bright (subsaturation) artificial satellite trail and several orders of crosstalk. The exact pattern and amplitude of channel-to-channel crosstalk varies among CCDs and readout electronics, but we find that it is stable and confined to a given CCD. Inter-CCD crosstalk is below the measurement error. Correction therefore requires measurement of the crosstalk coupling between each pair of 16 channels for each CCD. Crosstalk matrix measurement and correction is described in Section~\ref{subsec:crosstalk} and demonstrated in the bottom panel of Figure~\ref{fig:xtalk}, which shows the same trail after a preliminary nonlinear crosstalk correction method has been applied to the raw image.

Based on the hypothesis that the crosstalk is due to capacitive coupling between nearby signal lines, one would naively expect only positive crosstalk to be seen and only in the nearest-neighbor channels. However, we observe the crosstalk from trails to be bipolar and also nonlinear above some modest flux level. We know there are at least three sources of crosstalk: on-chip effects, ribbon cable capacitive coupling of video signals, and crosstalk originating in the electronics \citep{Antilogus_2017}. Further studies into the sources of the crosstalk and its nonlinearity are ongoing, and it may be possible to reduce some crosstalk effects at the hardware level. Regardless of the source, the crosstalk coefficients must be well characterized in intensity to allow correction at any brightness level.

\subsection{Non-linear crosstalk removal algorithm}
\label{subsec:crosstalk}
The crosstalk of the satellite trail presents a new challenge for image-correction algorithms because now the main trail has a variable multiplicative effect on neighboring channels, depending in a nonlinear way on the amplitude of the main trail. The correction algorithm we report here is preliminary and may be improved in both accuracy and speed in the future.

Simply masking the affected pixels would impact survey efficiency and uniformity, but doing so can introduce systematic errors. The crosstalk removal algorithm must specifically address these long, highly correlated crosstalk electronic ghost images at multiple positions over the affected CCD. These images could masquerade as faint sources or transient objects, as well as generate systematic errors via correlated lines of noise.

Crosstalk between the 16 video channels of our CCDs has been studied earlier \citep{O_Connor_2015}, but these measurements were limited and the CCD readout sequence has also been considerably changed. In our slit-illumination experiment, we observed nonlinear crosstalk between segments of the CCD at levels of $\sim5 \times 10^{-4}$ of trail flux, depending on the segment, but also nonlinearity in the crosstalk coefficients as a function of flux.

\begin{figure}[ht!]
\includegraphics[width=0.48\textwidth]{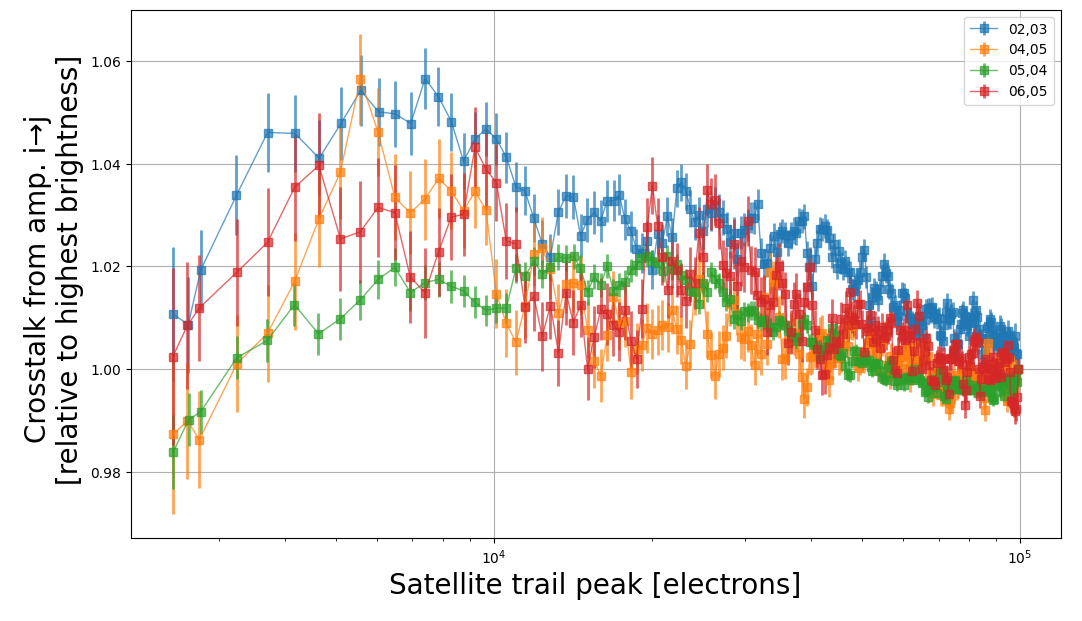}
\caption{The nonlinear dependence of crosstalk coefficients between nearest-neighbor channels of a single LSSTCam CCD, normalized to the coefficient at 100,000 $e^-$ per pixel.  Linear crosstalk, from capacitive coupling, would result in flat curves. Bench-top probes of the electronics have confirmed that the observed nonlinear behavior is likely the result of multiple competing sources in the readout chain. Further study is needed to confirm whether these variations can be sufficiently characterized for all pairs of 16 channels on each of the 189 CCDs in the LSSTCam such that satellite trail crosstalk artifacts are not a limiting systematic error in the survey. \label{fig:relxtalk}}
\end{figure}

Figure~\ref{fig:relxtalk} shows the nonlinear behavior we  measured on one of the LSSTCam e2v CCDs in the laboratory at the University of California Davis, showing crosstalk versus satellite trail illumination for various nearest-neighbor segments of the CCD. The measurements are reproducible and stable within the errors shown. If crosstalk were linear, all curves in Figure~\ref{fig:relxtalk} would be flat at 1.0. Instead, the crosstalk in nearest-neighbor channels is measurably nonlinear, with variations in crosstalk coefficients of 10\% or larger across the full response of the CCD.
There are nonlinear nonmonotonic deviations in crosstalk. Channels that are two or more apart show an even larger nonlinear dependence, although the crosstalk amplitude itself is smaller and sometimes negative.

To measure crosstalk, we take the ratio of overscan (bias level) subtracted pixel values of the ``crosstalk trail" channel to the ``main trail" channel. The former has been corrected for the small scattered light background that the crosstalk signal is superimposed upon, and the latter has had the bias level of the CCD subtracted. Estimating the bias and background levels is critical to the crosstalk measurement process as it provides the baseline upon which crosstalk is superimposed, and errors in its estimation can introduce systematic errors in the crosstalk coefficients used for correction.

At each intensity level, these crosstalk matrix coefficients are measured between all 16 channels in a CCD, forming the basis of a $16\times16$ nonlinear crosstalk matrix. The ratio is calculated for each of thousands of pixels in each channel image at each intensity to form the measurement of crosstalk. These flux-dependent crosstalk coefficients are plotted in Figure~\ref{fig:relxtalk}, where the points and error bars represent the mean and standard deviation of the distribution of crosstalk ratios measured between pairs of neighboring channels shown in differing colors. The error bars in the figure represent the combination of statistical error due to Poisson counts and errors in the estimation of the background level.

Crosstalk correction is finally performed by multiplying the nonlinear crosstalk coefficients by each pixel of the main trail channel (which has been bias overscan corrected), and this product is subtracted from raw crosstalk trail channels. Preliminary tests of this measurement and correction method have been successful on the $f/1.2$ beam simulator trails as shown in Figure~\ref{fig:xtalk}, as well as on LSSTCam hardware, which exhibits similar nonlinear crosstalk behavior.

Crosstalk between subsections of a CCD that are read out simultaneously has been known for some time. Indeed, most CCD cameras exhibit some level of video channel crosstalk. What is new is the throughput of the LSSTCam, the nonlinearity of its crosstalk response, and the sensitivity of the full survey to low-level systematics. For a given crosstalk measurement precision during operations, there is a maximum satellite trail brightness for which the crosstalk trails can be reduced to a small fraction of sky noise. Our measurements of the CCD impact of the satellite trails indicate that trails with more than {about 5000--10,000 $e^-$ induce} disproportionately large and variable crosstalk relative to their parent streak, corresponding to LEOsats brighter than approximately seventh apparent $g$ magnitude based on the LSSTCam exposure simulations described in Section~\ref{sec:sim-cam-response}.
Providing that the peak trail flux is lower than {about 5000--10,000 $e^-$, depending on the precision of measurement, the} preliminary nonlinear correction algorithm we report here can correct for most of the variation in crosstalk coefficients, assuming that they are stable during the survey and can be characterized to the precision presented here. However, further study is still needed to determine the origin of the crosstalk and improve its correction in hardware and implementation in software, as well as to validate that it can correct the full dynamic range of each satellite trail down to {well below} the noise level of the survey.
{The issue of measurement precision is illustrated by the two gray bands in Figures~\ref{fig:peakesim550} and \ref{fig:peaksim1200}. If the net measurement plus correction precision of the small crosstalk elements is 20\%, then the light gray band shows the range of correction to 10\% of the sky noise.  The dark band corresponds to a net crosstalk measurement plus correction precision of 10\%.}


\section{Discussion}
\label{sec:discuss}

Our motivation is to evaluate the impact and investigate solutions to mitigate the optical interference created by any LEOsat constellation. The SpaceX Starlink constellation is the first to deploy many hundreds of LEOsats, and the company engaged with the astronomical community to share design and operational information. They also tested and quickly fielded experiments to explore mitigation of Starlink visibility. We are therefore able to analyze the impact of actual LEOsats at various operational phases and undertake mitigation studies. While Rubin Observatory is the present limiting case because of its unprecedented etendue (but we recall that observations with very long exposure times are very vulnerable as well), most other observatories---ground-based and space-based---{will be affected indirectly because some of their science programs over the coming decade will rely on LSST data in place of the Sloan Digital Sky Survey. This is particularly true for transient follow-up.} Moreover, other LEOsat operators in the coming decade will benefit from this mitigation work in their efforts to be environmentally friendly.

Combining the analyses in Sections~\ref{sec:sim-cam-response} and \ref{sec:lab-trail-sim}, to get well into the linear response region for LSSTCam, LEOsats should not be brighter than $g\sim7$ mag at any airmass. Our analysis of the DECam data in Section~\ref{subsec:decam-photometry} shows that v1.0 Starlink satellites are brighter than this---closer to $g\sim5.1$ (no mitigation) and $g\sim6.1$ (DarkSat) at zenith.
Making the connection to physical radiant intensity for the satellite due to reflected sunlight, the seventh apparent magnitude for a satellite viewed at a range of 1000 km is equivalent to 44 W sr$^{-1}$.

At the time of this writing, the SpaceX effort to darken Starlink spacecraft, with both the DarkSat and VisorSat experiments, is on track to reach the level where we think we can suppress most or all LSSTCam artifacts from the resulting fainter satellite trail.
This is a promising development, but after suppressing the artifacts, we are left with the satellite trails themselves. While it has not been decided exactly how the LSST Project will handle satellite trails, they are likely to be masked in the data products, much as saturated pixels from bright stars will be masked. The LSST Project will do what is expedient, optimized for the general user community. Whatever the LSST Project ultimately does to remove the trails from the catalog, some signature of that process will remain, and the science data analysis may be variously sensitive to such signatures. The fraction of lost pixels is small but not zero. The most significant science impact may arise from systematic errors caused by low surface brightness residuals from the processing of satellite trails. The science community may have to do some amount of extra work to reach the promise of using the LSST to discover the unexpected. There may be cost and schedule impacts, and the presence of LEOsats may require the LSST to run for longer than 10 years to achieve all science goals.

In addition to the visibility of Starlink satellites when on-orbit, where they are expected to operate for 5--7 yr, the astronomy community has also noted the impact from the trains of multiple LEOsats in the 4--8 weeks following deployment, when they are operating at a lower-altitude parking orbit at 380 km, before they are raised to 550 km.
These Starlink satellites can appear many magnitudes brighter due to the ``open book'' configuration of the solar panel in this operational phase, where solar panel and satellite bus are coplanar and aligned with the velocity vector in order to reduce drag. In this configuration, Starlink satellites have been reported at 1--2 $g$ mag, with flares to -2 mag~\citep{2020AAS...23541003S}. While this operational phase is significantly shorter in duration than the on-orbit phase, SpaceX has been maintaining a regular cadence of Starlink launches, each deploying 60 satellites
in order to populate the constellation for useful broadband service and to meet U.S. and international regulatory deployment milestones.
SpaceX estimates 200--300 such satellites will be deployed in this steady state during their active deployment periods. In order to mitigate the significant brightness of Starlink satellites during these shorter periods, SpaceX is employing an operational mitigation of rolling the satellite bus edge-on to the sun to reduce the projected area illuminated by the sun, and diffuse reflections visible from the ground. This operational-roll technique was first tested in 2020 April, along with observations to determine its effectiveness. It is estimated that along with accurate orbit information provided publicly by constellation operators such as SpaceX, Rubin Observatory will be able to avoid as many as 300 known bright objects such as LEOsats in an optimized observation scheduler.

Taking multiple exposures is a partial mitigation. When the nominal LSST visit time of 30 s is split into two back-to-back exposures of 15 s, as currently planned, the comparison of these exposures using difference imaging could be used to identify a satellite trail. The exposure with the satellite trail in it can be rejected,  or the trail can be masked. This mitigation scenario would cost 8\% of the LSST observing time in order to accommodate the additional readout time and shutter motion, assuming a negligible cost due to rejected pixels, and it only mitigates some science.

Ultimately, we should plan on a combination of {the best of} these mitigation measures.


\section{Remaining challenges and plans}
\label{sec:challenges}

If each LEOsat can be darkened to approximately seventh $g$ mag during Rubin Observatory operations, we may be able to correct for the many image artifacts caused by satellite trails at this level, and most science may be unaffected. However, this
conclusion relies on fewer than $\sim48,000$ LEOsats in approximately 500--600 km orbits, as well as all satellite operators darkening their LEOsats to seventh $g$ mag or fainter.
LEOsats at 1200 km present another challenge because at this altitude, they are visible all night long~\citep{2020AAS...23541003S}.

We have no way to guarantee other LEOsat companies will follow the darkening example set by SpaceX, and no way to know how many LEOsats will ultimately be present during LSST operations. Rubin Observatory and SpaceX are committed to continuing their joint effort in assessing both the impact of Starlink LEOsats and the effectiveness of mitigation techniques as identified, fielded, and observed. We plan to revisit this analysis in another paper after the next iteration of signature reduction (VisorSat) reaches operational altitude, photometric observations are completed and analyzed, more progress is made on image artifact suppression, and after exploration of new dodging algorithms. The conclusions of this paper are predicated on all future LEOsats having successful on-satellite darkening mitigations to the $g\sim7$ AB mag level.

Some LSST science is particularly sensitive to low-level systematic errors. Other transient object science can be affected by the trails left by LEOsats, even with mitigations. Additional impacts arise from the processing, detection, cataloging, and science analysis overheads due to any satellite trails. Even with LEOsats darkened to seventh magnitude, satellite trails will still exist at the level of $\sim100$ times sky background noise. These trails will generate systematic errors that may impact data analysis and limit some science.
It remains to be seen if it will be feasible to custom-model and subtract each trail to high precision.

In the past, sky survey science has been limited by sample size $n$, so that statistical root-$n$ errors have dominated. With the unprecedented 40 billion objects expected from the LSST, the situation is different. LSST science will be mostly limited by systematic errors, and model-subtracted or masked satellite trails will contribute to the systematic error budget, along with bright stars and other masked sources.
However, the imprint of these two types of masks has different types of symmetry: stars have point symmetry, and trails have line symmetry. Some measures of cosmology are symmetry-dependent and may be affected by these kinds of systematic errors at low surface brightness.  It is useful to compare the expected satellite trail brightness with the faint limits the LSST is expected to reach.
For example, a relatively faint 10,000 $e^-$ per pixel LEOsat trail would have a surface brightness about 1000 times greater than most galaxies in the LSST. By comparison, one of the faint galaxies in our ``gold sample'' of several billion galaxies has $\sim$12 $e^-$ per pixel average surface brightness in a 30 s $g$-band exposure (equivalent to 26.5 $g$ mag arcsec$^{-2}$). To avoid obvious residuals, the process of satellite trail removal from the LSST alerts and database would have to achieve a surface brightness precision of 3e--3 on each exposure. As shown in Figure~\ref{fig:widthsim550}, this would require special masking scaled to each trail.
\added{For example, the bright time sky noise in a 30 s exposure in $r$ band is about 85 $e^-$ per pixel, and a $5\sigma$ threshold for the automated detect/mask gives 400 $e^-$ per pixel at the edge of the initial mask. This would be diluted to about 4 $e^-$ per pixel in the coadd. Because the faint galaxies used in lensing have outer surface brightness below 1 $e^-$ per pixel, we instead would grow the mask in an exposure so that the contribution to the coadd at the grown mask edge is less than this --- leading to the conservative grown mask reaching a level of 1\% sky noise in an exposure.}

There are eight Rubin Observatory science collaborations: Galaxies; Stars, Milky Way, and Local Volume; Solar System; Dark Energy; Active Galactic Nuclei; Transients and Variable Stars; Strong Lensing; and Information and Statistics.
The LSST Science Book \citep{scibook}\footnote{\url{https://www.lsst.org/scientists/scibook}} outlines over 100 examples of unprecedented science reach in many types of probes of our universe using LSST data. Of course we cannot explicitly list the unexpected discoveries; LSST is specifically designed to search for the unexpected, and many of the same characteristics that make LSST vulnerable to LEOsats also make it ideally suited for this.

Using this first study of the possible impacts of LEOsat trails on the LSST data, it will be possible over the coming year for each science collaboration to undertake simulations of impact on their particular science programs.
We need to investigate bogus signals or systematic errors resulting from LEOsat artifacts in the images and catalog and the degree to which they might negatively affect LSST science programs. As one example, the lines of correlated pixels due to residuals after trail removal could bias weak gravitational lensing cosmic shear probes of the nature of dark energy and dark matter. To investigate the level of cosmic shear noise arising from the cumulative effects of this small print-through bias, future work should include full simulations of the LSST (spanning $\sim$20,000 deg$^2$ and with 50--150 visits per field per filter band) with many long stripes of no data to simulate satellite trails.

Another example is the impact on transient and moving-object detections. Specifically, bogus transient events and false alerts, as well as tracklet linkage degradation in planetary programs, especially those designed to detect near-Earth asteroids in early twilight. Taken together, these are only a handful of examples of scientific impact of the LEOsat trail mitigation that the scientific community needs to investigate. This represents significant work beyond the original scope needed to do science with the Rubin Observatory LSST, and will slow the pace of discovery and scientific advancement.

We plan a second paper on this subject in the next year to report on the next phase of SpaceX mitigation experiments as well as preliminary science impact simulations. The Rubin Observatory commissioning camera (one 36 megapixel $3\times3$ CCD raft installed on the telescope in advance of the main 3.2 gigapixel LSSTCam) will be the first-light instrument. Direct tests of the effects of LEOsat trails on the LSST will be a natural part of the commissioning camera's mission of validating the telescope and observatory operations via its on-sky observing campaign.

\acknowledgments

We thank Todd Boroson, Polina Danilyuk, Jonathan Herman, Craig Lage, Josh Meyers, and Perry Gee for help. We also thank Patricia Cooper, Alex Drlica-Wagner, David Goldman, Steve Heathcote, Robert Lupton, Phil Puxley, Steve Ritz, Aaron Roodman, Clare Saunders, Pat Seitzer, Adam Snyder, and Chris Stubbs for useful discussions. M.L.R. thanks Chris Waters for assisting with DECam image analysis.

This research was supported by the Department of Energy (DOE) grant DE-SC0009999, National Science Foundation and Association of Universities for Research in Astronomy (NSF/AURA) grant N56981C, and NSF grant AST-2024216. This work was initiated at the Aspen Center for Physics, which is supported by NSF grant PHY-1607611.

The Rubin Observatory project is jointly funded by the National Science Foundation (NSF) and the Department of Energy (DOE) Office of Science, with early construction funding received from private donations through the LSST Corporation. The NSF-funded Project Office for construction was established as an operating center under management of the Association of Universities for Research in Astronomy (AURA). The DOE-funded effort to build the Rubin Observatory LSST Camera (LSSTCam) is managed by the SLAC National Accelerator Laboratory (SLAC).

NSF and DOE will continue to support Rubin Observatory in its Operations phase to carry out the Legacy Survey of Space and Time. They will also provide support for scientific research with the data. During operations NSF funding is managed by the Association of Universities for Research in Astronomy (AURA) under a cooperative agreement with NSF, and DOE funding is managed by SLAC under contract by DOE. The Vera C. Rubin Observatory is operated by NSF's Optical-Infrared Astronomy Research Laboratory and SLAC.

This project used data obtained with the Dark Energy Camera (DECam), which was constructed by the Dark Energy Survey (DES) collaboration. Funding for the DES Projects has been provided by the US Department of Energy, the US National Science Foundation, the Ministry of Science and Education of Spain, the Science and Technology Facilities Council of the United Kingdom, the Higher Education Funding Council for England, the National Center for Supercomputing Applications at the University of Illinois at Urbana-Champaign, the Kavli Institute for Cosmological Physics at the University of Chicago, Center for Cosmology and Astro-Particle Physics at the Ohio State University, the Mitchell Institute for Fundamental Physics and Astronomy at Texas A\&M University, Financiadora de Estudos e Projetos, Fundação Carlos Chagas Filho de Amparo à Pesquisa do Estado do Rio de Janeiro, Conselho Nacional de Desenvolvimento Científico e Tecnológico and the Ministério da Ciência, Tecnologia e Inovação, the Deutsche Forschungsgemeinschaft and the Collaborating Institutions in the Dark Energy Survey.

The Collaborating Institutions are Argonne National Laboratory, the University of California at Santa Cruz, the University of Cambridge, Centro de Investigaciones Enérgeticas, Medioambientales y Tecnológicas–Madrid, the University of Chicago, University College London, the DES-Brazil Consortium, the University of Edinburgh, the Eidgenössische Technische Hochschule (ETH) Zürich, Fermi National Accelerator Laboratory, the University of Illinois at Urbana-Champaign, the Institut de Ciències de l’Espai (IEEC/CSIC), the Institut de Física d’Altes Energies, Lawrence Berkeley National Laboratory, the Ludwig-Maximilians Universität München and the associated Excellence Cluster Universe, the University of Michigan, NSF’s NOIRLab, the University of Nottingham, the Ohio State University, the OzDES Membership Consortium, the University of Pennsylvania, the University of Portsmouth, SLAC National Accelerator Laboratory, Stanford University, the University of Sussex, and Texas A\&M University.

Based on observations at Cerro Tololo Inter-American Observatory, NSF’s NOIRLab (NOIRLab Prop. ID 2019A-0305; PI: A. Drlica-Wagner), which is managed by the Association of Universities for Research in Astronomy (AURA) under a cooperative agreement with the National Science Foundation.

\facility{Blanco (DECam).}

\software{astropy \citep{astropy:2013, astropy:2018},
          SciPy \citep{scipy},
          Numpy \citep{numpy},
          Pandas \citep{pandas},
          Matplotlib \citep{matplotlib},
          LSST Science Pipelines \citep{2019ASPC..523..521B},
          GalSim \citep{galsim},
          LSST Simulations Photometric Utilities (\url{https://github.com/lsst/sims_photUtils}),
          LSST Project Science Team System Engineering Throughputs (\url{https://github.com/lsst-pst/syseng_throughputs})
          }


\bibliography{LEOsat-mitigation}{}
\bibliographystyle{aasjournal}

\end{document}